\documentclass[twocolumn,superscriptaddress,letter]{revtex4}%
\usepackage{amssymb}
\usepackage{color}
\usepackage{graphicx}
\usepackage{dcolumn}
\usepackage{bm}
\usepackage[header,title,page,titletoc]{appendix}
\usepackage{amsmath}
\usepackage{amsfonts}%
\providecommand{\U}[1]{\protect \rule{.1in}{.1in}}
\begin{document}
\title{Weyl Bogoliubov excitations in the Bose-Hubbard extension of a Weyl semimetal}
\author{Ya-Jie Wu}
\affiliation{School of Science, Xi'an Technological University, Xi'an 710032, China}
\author{Wen-Yan Zhou}
\affiliation{Center for Advanced Quantum Studies, Department of Physics, Beijing Normal
University, Beijing, 100875, China}
\author{Su-Peng Kou}
\email{spkou@bnu.edu.cn}
\affiliation{Center for Advanced Quantum Studies, Department of Physics, Beijing Normal
University, Beijing, 100875, China}

\begin{abstract}
In this paper, a Bose-Hubbard extension of a Weyl semimetal is proposed that
can be realized for ultracold atoms using laser assisted tunneling and
Feshbach resonance technique in three dimensional optical lattices. The global
phase diagram is obtained consisting of a superfluid phase and various Mott
insulator phases by using Landau theory. The Bogoliubov excitation modes for
the weakly interacting case have nontrivial properties (Weyl nodes, bosonic
surface arc, etc.) analogs of those in Weyl semimetals of electronic systems,
which are smoothly carried over to that of Bloch bands for the noninteracting
case. The properties of the insulating phases for the strongly interacting
case are explored by calculating both the quasiparticle and quasihole
dispersion relation, which shows two quasiparticle spectra touch at Weyl nodes.

\bigskip \noindent PACS number(s): 37.10.JK, 03.75.Fi, 67.40.-w, 32.80.Pj

\end{abstract}
\maketitle

%\email{wuyajie@xatu.edu.cn}

\section{Introduction and motivation}

Recently, Weyl semimetal (WSM) attracts considerable interest in both theory
and experiments \cite{wan,lu,tur,lv1,lv2,xu,xu1,xu2,sol,li,kong}. Differently
from gapped topological insulators and superconductors, WSM has bulk gapless
nodal points (dubbed Weyl points), which exhibit topological structure of
synthetic monopole in momentum space and give rise to surface Fermi arc that
connects different chiral Weyl points. To realize Weyl points, time-reversal
and/or inversion symmetry of the system must be broken \cite{tur,lu}.
According to the emergent Lorentz invariance of Weyl points, WSMs are
classified into type-I with Lorentz-invariance-preserving Weyl nodes
\cite{wan,lu,tur,lv1,lv2,xu}, type-II with Lorentz-invariance-violating Weyl
nodes \cite{xu1,xu2,sol}, and hybrid WSM with mixed types of Weyl nodes
\cite{li,kong}.

In addition, the investigation of topology of bosonic modes attracts
increasing attention in interacting bosons in optical lattices \cite{Fur,enge}%
, photonic systems \cite{vitt,vitt1}, magnonic excitations
\cite{shi1,shi2,fei}, phononic excitations \cite{pro,sus} and polaritonic
excitations \cite{bar,kar}, etc. In general, the bosons condense into the mode
with lowest energy at zero temperature. However, the energy bands of excited
bosonic modes may exhibit topological structure
\cite{Fur,enge,shi1,shi2,fei,pro,sus,bar,kar,sta,wong}, which gives rise to
topologically protected edge modes in the excitation spectrum owing to the
bulk-boundary relation \cite{Fur}.

Rapid progress on synthetic magnetic and gauge fields in ultracold atoms
provides opportunities for realizing novel states of matter
\cite{dal,gold,car,zhan}. By using laser-assisted tunneling in three
dimensional optical lattices, Tena Dub\v{c}ek et al. proposed that WSM with
broken inversion symmetry may be realized in optical lattices \cite{ten}.
Alongside advances in manipulating ultracold atoms and novel properties of
WSM, an interesting issue arises: \textquotedblleft what are the properties of
excitation modes in superfluid phase and Mott-insulator phases in Bose-Hubbard
extension of the Hamiltonian with Weyl points?\textquotedblright \ Therefore,
in this paper we focus on the Bogoliubov excitations of the Bose-Hubbard
extension of WSM. It is found that the energy dispersion of Bogoliubov modes
exhibits Weyl points with chirality and there exist topologically protected
bosonic surface-arc states in both superfluid phase and Mott-insulator phases
analogs of that in WSMs of electronic systems.

The remainder of the paper is organized as follows. In Sec. \ref{sec2}, we
first present the Bose-Hubbard extension of the WSM, and then give the band
structure for its noninteracting case. By means of Landau theory, we show the
phase diagram that consists of superfluid phase and Mott-insulator phases. In
Sec. \ref{sec3}, we calculate the Bogoliubov excitation bands for bosonic
superfluids by using Bogoliubov theory, and present that there are Weyl points
analogs of that in WSMs of electronic systems. The bosonic surface-arc states
on boundaries that are connected by Weyl points with different chiralities are
found. In Sec. \ref{sec4}, by using the functional integral formalism, we
derive the quasiparticle and quasihole dispersions in Mott-insulator phase
that show two quasiparticle spectra touch at Weyl points. Finally, we conclude
our discussions in Sec. \ref{sec5}.

\section{Bose-Hubbard extension of the Weyl semimetal}

\label{sec2}

The Hamiltonian of the Bose-Hubbard extension of the WSM in three-dimensional
lattices is given by%
\begin{equation}
\hat{H}=\hat{H}_{0}+\frac{U}{2}\sum_{\mathbf{r}}\hat{a}_{\mathbf{r}}%
^{\dagger2}\hat{a}_{\mathbf{r}}^{2}-\mu \sum_{\mathbf{r}}\hat{a}_{\mathbf{r}%
}^{\dagger}\hat{a}_{\mathbf{r}}.
\end{equation}
Here, $U$ is the on-site interaction strength, $\mu$ is the chemical
potential, and the Weyl Hamiltonian $H_{0}$ takes the following form as
\cite{ten}
\begin{align}
\hat{H}_{0} &  =\sum_{\mathbf{r\in B}}\left(  -J_{x}\hat{a}_{\mathbf{r}%
+\mathbf{\delta}_{x}}^{\dagger}\hat{a}_{\mathbf{r}}+J_{x}\hat{a}%
_{\mathbf{r}-\mathbf{\delta}_{x}}^{\dagger}a_{\mathbf{r}}-J_{y}\hat
{a}_{\mathbf{r}+\mathbf{\delta}_{y}}^{\dagger}\hat{a}_{\mathbf{r}}%
-J_{y}\right.  \nonumber \\
&  \left.  \times \hat{a}_{\mathbf{r}-\mathbf{\delta}_{y}}^{\dagger}\hat
{a}_{\mathbf{r}}+h.c.\right)  +\sum_{\mathbf{r\in}A}J_{z}\hat{a}%
_{\mathbf{r}+\mathbf{\delta}z}^{\dagger}\hat{a}_{\mathbf{r}}-\sum
_{\mathbf{r\in}B}J_{z}\hat{a}_{\mathbf{r}+\mathbf{\delta}_{z}}^{\dagger}%
\hat{a}_{\mathbf{r}}\label{eq0}%
\end{align}
where $\hat{a}_{\mathbf{r}}$ denotes the bosonic annihilation operator at the
lattice site $\mathbf{r}$ ($\mathbf{r\in}A$\textbf{, }$B$
sublattices\textbf{)}, $J_{x}$, $J_{y}$ and $J_{z}$ are the real
nearest-neighbor hopping parameters along the $x$, $y$, and $z$ direction,
respectively. Here, we introduce vectors $\mathbf{\delta}_{1,2,3}$ given by
$\mathbf{\delta}_{x}=d\left(  1,0,0\right)  ,\mathbf{\delta}_{y}=d\left(
0,1,0\right)  ,\mathbf{\delta}_{z}=d\left(  0,0,1\right)  $, where $d$ is the
length between the neighboring sites. 

\subsection{Bloch band structure for the Weyl semimetal}

Firstly, we briefly review the band structure of the Hamiltonian with Weyl
points in the noninteracting case ($U=0$). Provided that the system has
periodic boundary condition, we perform the Fourier transformation $\hat
{a}_{\mathbf{r}}=\frac{1}{\sqrt{N_{uc}}}\sum_{\mathbf{k}}\hat{a}%
_{X,\mathbf{k}}e^{i\mathbf{k}.\mathbf{r}}$, where $\mathbf{r}\in X=A,B$, the
sum is taken over the discrete momenta\ $\mathbf{k}$ in the first Brillouin
zone, and $N_{uc}$ is the number of unit cells in the system. The Hamiltonian
$\hat{H}_{0}$ in momentum space is then given by%
\begin{equation}
\hat{H}_{0}=\sum_{\mathbf{k}}\left(  \hat{a}_{A,\mathbf{k}}^{\dagger},\hat
{a}_{B,\mathbf{k}}^{\dagger}\right)  \mathcal{H}\left(  \mathbf{k}\right)
\left(
\begin{array}
[c]{c}%
\hat{a}_{A,\mathbf{k}}\\
\hat{a}_{B,\mathbf{k}}%
\end{array}
\right)  .\label{eq1}%
\end{equation}
Here, the Hermitian matrix $\mathcal{H}\left(  \mathbf{k}\right)  $ is
described by%
\begin{equation}
\mathcal{H}\left(  \mathbf{k}\right)  =\vec{h}\left(  \mathbf{k}\right)
\cdot \vec{\sigma},\label{eq2}%
\end{equation}
where $I$ is the identity matrix, $\vec{\sigma}=\left(  \sigma_{1},\sigma
_{2},\sigma_{3}\right)  $ are Pauli matrices, and the coefficients $\vec
{h}\left(  \mathbf{k}\right)  =\left(  h_{1}\left(  \mathbf{k}\right)
,h_{2}\left(  \mathbf{k}\right)  ,h_{3}\left(  \mathbf{k}\right)  \right)  $
are written as%
\begin{align}
h_{1}\left(  \mathbf{k}\right)   &  =-2J_{y}\cos \left(  k_{y}d\right)
,h_{2}\left(  \mathbf{k}\right)  =-2J_{x}\sin \left(  k_{x}d\right)
,\nonumber \\
h_{3}\left(  \mathbf{k}\right)   &  =2J_{z}\cos \left(  k_{z}d\right)  .
\end{align}
In the follwing, we set $d=1$. The two energy bands are obtained through the
diagonalization of Eq. (\ref{eq1}) as%
\begin{equation}
e_{\pm}\left(  \mathbf{k}\right)  =\pm h\left(  \mathbf{k}\right)  =\pm
\sqrt{h_{1}^{2}\left(  \mathbf{k}\right)  +h_{2}^{2}\left(  \mathbf{k}\right)
+h_{3}^{2}\left(  \mathbf{k}\right)  }.
\end{equation}
Hereafter, we choose $J_{x}=J_{y}=J_{z}=J$. The system then has four
independent Weyl points at $\left \{  \mathbf{k}_{w}\right \}  =\left(  0,\pm
\pi/2,\pm \pi/2\right)  $ \cite{ten}. For non-interacting bosons, Bose-Einstein
condensation (BEC) into the lowest-energy single-particle states occurs at
zero temperature. The bottom of the lowest band $e_{-}\left(  \mathbf{k}%
\right)  $ is located at four different momenta of the Brillouin zone
$\left \{  \mathbf{k}_{0}\right \}  =\left(  \pm \pi/2,0,0\right)  $, $\left(
\pm \pi/2,0,\pi \right)  $. Here, we consider that the BEC is prepared at
$\mathbf{k}_{0}=\left(  \pi/2,0,0\right)  $. At this time, the coefficients
$\vec{h}\left(  \mathbf{k}_{0}\right)  $ are given by $\left(  -2J_{y}%
,-2J_{x},2J_{z}\right)  $.

To determine the single-particle ground state, we parameterize $\vec{h}\left(
\mathbf{k}_{0}\right)  $ using the spherical coordinate as%
\begin{equation}
\vec{h}\left(  \mathbf{k}_{0}\right)  =h\left(  \mathbf{k}_{0}\right)  \left(
\sin \theta_{0}\cos \varphi_{0},\sin \theta_{0}\sin \varphi_{0},\cos \theta
_{0}\right)
\end{equation}
with $h\left(  \mathbf{k}_{0}\right)  =2\sqrt{3}J$, and $\left(  \theta
_{0},\varphi_{0}\right)  =\left(  \arccos \left(  1/\sqrt{3}\right)
,\arccos \left[  -1/\left(  \sqrt{3}\sin \theta_{0}\right)  \right]  \right)  $.
Then the $\mathbf{k}_{0}$ part of the Hamiltonian $\hat{H}_{0}$ may be
diagonalized by using the transformation
\begin{equation}
\left(
\begin{array}
[c]{c}%
\hat{a}_{A,\mathbf{k}_{0}}\\
\hat{a}_{B,\mathbf{k}_{0}}%
\end{array}
\right)  =\mathcal{U}\left(  \theta_{0},\varphi_{0}\right)  \left(
\begin{array}
[c]{c}%
\hat{a}_{+,\mathbf{k}_{0}}\\
\hat{a}_{-,\mathbf{k}_{0}}%
\end{array}
\right)  , \label{eqa0}%
\end{equation}
where the unitary matrix $\mathcal{U}\left(  \theta_{0},\varphi_{0}\right)  $
is given by
\begin{equation}
\mathcal{U}\left(  \theta_{0},\varphi_{0}\right)  =\left(
\begin{array}
[c]{cc}%
e^{-i\varphi_{0}}\cos \left(  \frac{\theta_{0}}{2}\right)  & -e^{-i\varphi_{0}%
}\sin \left(  \frac{\theta_{0}}{2}\right) \\
\sin \left(  \frac{\theta_{0}}{2}\right)  & \cos \left(  \frac{\theta_{0}}%
{2}\right)
\end{array}
\right)  .
\end{equation}
For noninteracting bosons, BEC occurs in the mode created by $\hat
{a}_{-,\mathbf{k}_{0}}$. For interacting bosons, the condensate wave function
will be modified as the interaction gradually increases.

\begin{figure}[ptb]
\scalebox{0.30}{\includegraphics* [0.0in,0.0in][12.0in,4.3in]{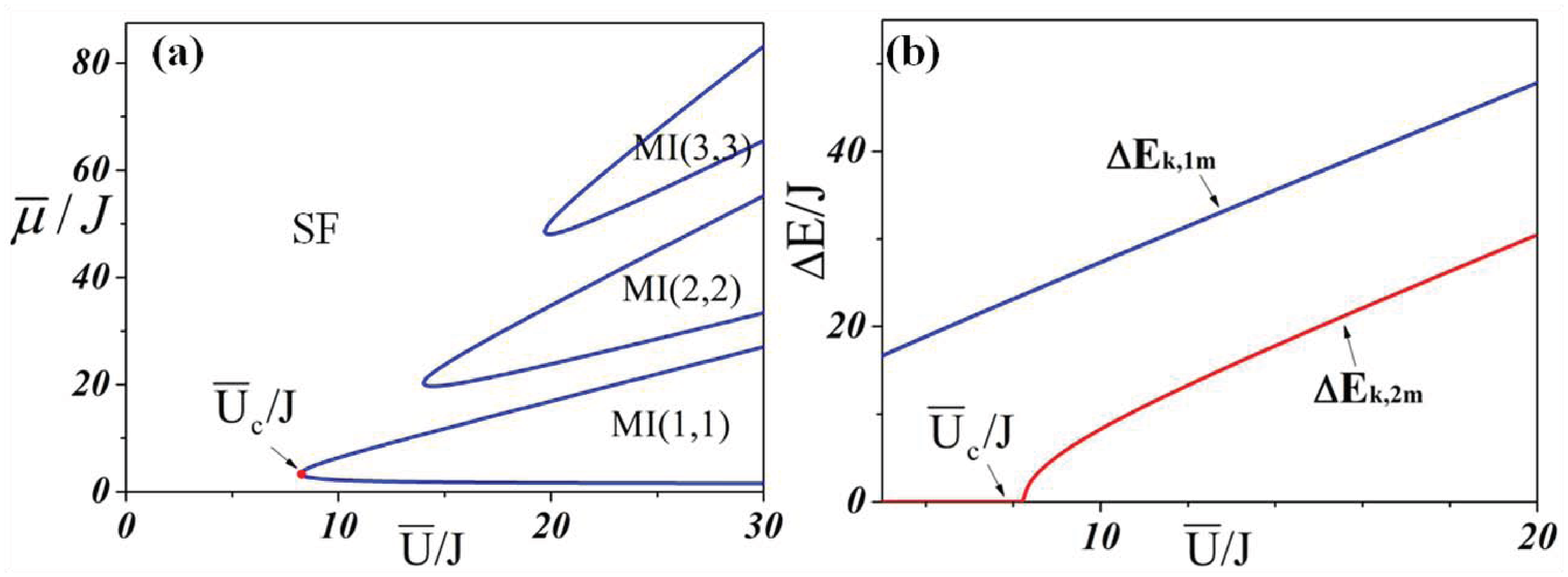}}\caption{(Color
online) (a) Phase diagram of the Bose-Hubbard extension of WSM. The vertical
axis and horizontal axis show the dimensionless chemical potential $\bar{\mu
}=\mu/z$ and $\bar{U}=U/z$, respectively. (b) The first-order approximations
to the dispersion of the density fluctuations.}%
\label{mi1}%
\end{figure}

\subsection{Superfluid-Mott insulator transition}

After turning on the interaction $U$ ($>0$) between particles, the ground
state of the system enters into superfluid (SF) phase with noninteger number
of bosons at each site at zero temperature. As interaction increases,
qualitatively the interaction between particles will drive the system into
Mott insulator (MI) phase if $U\gg t$, in which the moving for a particle from
one site to another is energetically unfavorable.

\bigskip In the strong coupling limit, we first introduce a local superfluid
order parameter that is written as \cite{van,lim}
\begin{equation}
\psi_{\mathbf{r}}=\langle \hat{a}_{\mathbf{r}}^{\dag}\rangle=\langle \hat
{a}_{\mathbf{r}}\rangle.
\end{equation}
Owing to that there are two kinds of lattice sites in the model, i.e.,
$A$-sublattice and $B$-sublattice, we define order parameters as
$\psi_{\mathbf{r\in}A}\equiv \psi_{A}$ and $\psi_{\mathbf{r\in}B}\equiv \psi
_{B}$, respectively. With the help of $\psi_{\mathbf{r}}$, the hopping terms
in Eq. (\ref{eq0}) can be decoupled, and in the occupation numbers basis we
readily arrive at the per unit cell ground state energy for the system up to
the second-order perturbation as%
\begin{align}
E_{g}\left(  \psi_{A},\psi_{B}\right)   &  \simeq a_{0}+a_{2}\psi_{A}%
^{2}+c_{2}\psi_{A}\psi_{B}\nonumber \\
&  +b_{2}\psi_{B}^{2}+\mathcal{O}\left(  \psi_{A}^{4},\psi_{B}^{4}\right)  .
\end{align}
Here, coefficients $a_{0}$, $a_{2}$, $c_{2}$, and $b_{2}$ are as follows:
\begin{equation}
a_{0}=\frac{\overline{U}}{2}(n_{A}^{2}+n_{B}^{2}-n_{u})-\overline{\mu}n_{u},
\label{eqq1}%
\end{equation}%
\begin{align}
a_{2}  &  =-J_{z}+\frac{(-\overline{U}-\mu)J_{z}{}^{2}}{\left[  (n_{A}%
-1)\overline{U}-\overline{\mu}\right]  \left[  -n_{A}\overline{U}%
+\overline{\mu}\right]  }\nonumber \\
&  +\frac{(-\overline{U}-\overline{\mu})(J_{x}+J_{y})^{2}}{\left[
(n_{B}-1)\overline{U}-\overline{\mu}\right]  \left[  -n_{B}\overline
{U}+\overline{\mu}\right]  },
\end{align}%
\begin{align}
b_{2}  &  =J_{z}+\frac{(-\overline{U}-\overline{\mu})(J_{x}-J_{y})^{2}%
}{\left[  (n_{A}-1)\overline{U}-\overline{\mu}\right]  \left[  -n_{B}%
\overline{U}+\overline{\mu}\right]  }\nonumber \\
&  +\frac{(-\overline{U}-\overline{\mu})J_{z}^{2}}{\left[  (n_{B}%
-1)\overline{U}-\overline{\mu}\right]  \left[  -n_{B}\overline{U}%
+\overline{\mu}\right]  },
\end{align}%
\begin{align}
c_{2}  &  =2J_{y}+\frac{2(-\overline{U}-\overline{\mu})(J_{x}-J_{y})J_{z}%
}{\left[  (n_{A}-1)\overline{U}-\overline{\mu}\right]  \left[  -n_{A}%
\overline{U}+\overline{\mu}\right]  },\nonumber \\
&  +\frac{2(-\overline{U}-\overline{\mu})(J_{x}+J_{y})J_{z}}{\left[
(n_{B}-1)\overline{U}-\overline{\mu}\right]  \left[  -n_{B}\overline
{U}+\overline{\mu}\right]  }, \label{eqq4}%
\end{align}
where $\bar{U}=U/z$, $\bar{\mu}=\bar{\mu}/z$ with $z=2$ being the number of
nearest-neighbor sites in one direction, the average particle number in one
unit cell\textit{ }$n_{u}=n_{A}+n_{B}$\textit{,} $n_{A}$ and $n_{B}$ are
particle number on $A$- and $B$-sublattice, respectively. See Appendix
\ref{Appendix_1} for detailed calculations. By means of Landau theory, based
on the geometry knowledge, when the Gaussian curvature of the energy-order
parameters surface is zero at the point $\psi_{A}=\psi_{B}=0$, the phase
transition occurs \cite{chenb}. Therefore, we can obtain a function of
phase-transition line readily by%

\begin{equation}
\frac{\partial^{2}E}{\partial \psi_{A}^{2}}|_{\left(  0,0\right)  }%
\frac{\partial^{2}E}{\partial \psi_{B}^{2}}|_{\left(  0,0\right)  }-\left(
\frac{\partial^{2}E}{\partial \psi_{A}\partial \psi_{B}}|_{\left(  0,0\right)
}\right)  ^{2}=0,
\end{equation}
which leads to the function of phase-transition line:%

\begin{equation}
4a_{2}b_{2}-c_{2}^{2}=0. \label{ptr}%
\end{equation}
Solving Eq. (\ref{ptr}), yields
\begin{align}
\overline{\mu}  &  =\frac{1}{2}\left[  -\sqrt{2}+(2\tilde{n}-1)\overline
{U}\right. \nonumber \\
&  \left.  \pm \sqrt{2-2\sqrt{2}\overline{U}-4\sqrt{2}\tilde{n}\overline
{U}+\overline{U}^{2}}\right]  \label{mu}%
\end{align}
with the particle number\textbf{ }$\tilde{n}\equiv n_{A}=n_{B}$.
Correspondingly, the point of smallest $\bar{U}$ (denoted by $\bar{U}_{c}$)
for each lobe is
\begin{equation}
\bar{U}_{c}=\sqrt{2}+2\sqrt{2}\tilde{n}+2\sqrt{2\tilde{n}+2\tilde{n}^{2}}.
\end{equation}

In conclusion, by applying the Landau theory of phase transitions that treats
the interactions exactly and the hopping terms as perturbation, we get the
phase diagram in Fig. \ref{mi1} (a). It shows that the phase transition occurs
at $\bar{U}_{c}/J\approx8.25$ for the MI $\left(  1,1\right)  $ lobe with the
particle number configuration $\left \{  n_{A}=1,n_{B}=1\right \}  $.

For the case of $U<U_{c}$, the system is in the SF phase. According to the
Hugenholtz-Pines theorem, there are always gapless density fluctuations. In
the followings, we will study the excitation modes in SF phase.

\section{Bogoliubov theory and topology of excitation modes for superfluid}

\label{sec3}

In this section, by using the Bogoliubov theory for homogeneous condensates
with weak repulsive interactions, we determine the band structure of
Bogoliubov excitations. Here, homogeneous\ case refers to the situation where
the system has the periodicity of the lattice. We then study the topology of
Bogoliubov excitations, and determine whether the excitations have novel properties.

By means of the Gross-Pitaevskii (GP) theory, we derive the condensate wave
function to formulate the Bogoliubov theory for the boson system. In the GP
theory, we first introduce the GP energy function $E$ by replacing $\left(
\hat{a}_{\mathbf{r}},\hat{a}_{\mathbf{r}}^{\dagger}\right)  $ by $\left(
\psi_{\mathbf{r}},\psi_{\mathbf{r}}^{\dagger}\right)  $ in the Hamiltonian in
Eq. (\ref{eq0}), and minimize it with respect to $\left(  \psi_{\mathbf{r}%
},\psi_{\mathbf{r}}^{\dagger}\right)  $ under the constraint $\sum
_{\mathbf{r}}\left \vert \psi_{\mathbf{r}}\right \vert ^{2}=N$. Since the
single-particle ground state is formed at $\mathbf{k}_{0}$, we first introduce
the following homogeneous ansatz for the interacting case: $\psi_{\mathbf{r}%
}=\frac{1}{\sqrt{N_{uc}}}\psi_{_{X}}$ ($X\in A,B$) with $N_{uc}$ being the
number of unit cells. Next, we introduce the chemical potential $\mu$ as a
Lagrange multiplier to satisfy the particle-number constraint. The functional
to be minimized is then given by%
\begin{align}
E-\mu N  &  =\left(  \psi_{_{A}}^{\ast},\psi_{_{B}}^{\ast}\right)  \left[
\mathcal{H}\left(  \mathbf{k}_{0}\right)  -\mu I\right]  \left(
\begin{array}
[c]{c}%
\psi_{_{A}}\\
\psi_{_{B}}%
\end{array}
\right) \nonumber \\
&  +\frac{U}{N_{uc}}\left(  \left \vert \psi_{_{A}}\right \vert ^{4}+\left \vert
\psi_{_{B}}\right \vert ^{4}\right)  .
\end{align}
Minimizing $E-\mu N$ with respect to $\psi_{_{X}}^{\ast}$ ($X\in A,B$) gives a
homogeneous version of the GP equations:%
\begin{equation}
\left[  \mathcal{H}\left(  \mathbf{k}_{0}\right)  -\mu I\right]  \left(
\begin{array}
[c]{c}%
\psi_{_{A}}\\
\psi_{_{B}}%
\end{array}
\right)  +\frac{U}{N_{uc}}\left(
\begin{array}
[c]{c}%
\psi_{_{A}}^{\ast}\psi_{_{A}}^{2}\\
\psi_{_{B}}^{\ast}\psi_{_{B}}^{2}%
\end{array}
\right)  =0. \label{eq0b}%
\end{equation}
Since the single-particle ground state is created by $a_{-}^{\dagger}\left(
\mathbf{k}_{0}\right)  $ in Eq. (\ref{eqa0}), it is convenient to parameterize
$\left(  \psi_{_{A}},\psi_{_{B}}\right)  ^{T}$ as%
\begin{equation}
\left(
\begin{array}
[c]{c}%
\psi_{_{A}}\\
\psi_{_{B}}%
\end{array}
\right)  =\sqrt{N}\left(
\begin{array}
[c]{c}%
f_{_{A}}\\
f_{_{B}}%
\end{array}
\right)  =\sqrt{N}\left(
\begin{array}
[c]{c}%
-e^{-i\varphi}\sin \left(  \frac{\theta}{2}\right) \\
\cos \left(  \frac{\theta}{2}\right)
\end{array}
\right)  ,
\end{equation}
where $\theta=\theta_{0}$ when $U=0$. Multiplying Eq. (\ref{eq0b}) by $\left(
f_{_{A}}^{\ast},f_{_{B}}^{\ast}\right)  $ or $\left(  -f_{_{B}}^{\ast}%
,f_{_{A}}^{\ast}\right)  $ from the left, we get\begin{widetext}
\begin{align}
-h\left(  \mathbf{k}_{0}\right)  \left[  \cos \theta_{0}\cos \theta+\sin
\theta_{0}\sin \theta \cos \left(  \varphi_{0}-\varphi \right)  \right]
+2Un\left(  \left \vert f_{_{A}}\right \vert ^{4}+\left \vert f_{_{B}}\right \vert
^{4}\right)   &  =\mu,\label{eq3a}\\
h\left(  \mathbf{k}_{0}\right)  \left(  \cos \theta_{0}\sin \theta \cos
\varphi-\sin \theta_{0}\cos \varphi_{0}\cos \theta \right)  -Un\cos \varphi
\sin \theta \cos \theta &  =0,\label{eq3b}\\
h\left(  \mathbf{k}_{0}\right)  \sin \theta_{0}\sin \varphi_{0}+\mu \sin
\theta \sin \varphi-Un\sin \theta \sin \varphi &  =0\label{eq3c}%
\end{align}
\end{widetext}where $n=N/\left(  2N_{uc}\right)  $.

We now discuss excitations from the condensate ground state by using the
Bogoliubov theory. Firstly, $\hat{a}_{\mathbf{r}}$ is decomposed into the
condensate and noncondensate parts with the help of Fourier transformation as%
\begin{equation}
\hat{a}_{\mathbf{r}}=\frac{1}{\sqrt{N_{uc}}}f_{X}\hat{a}_{-}+\tilde
{a}_{\mathbf{r}}, \label{eq4}%
\end{equation}
where $\tilde{a}_{\mathbf{r}}=\frac{1}{\sqrt{N_{uc}}}\left[  -\epsilon
_{X}f_{\bar{X}}^{\ast}\hat{a}_{+}+\sum_{\mathbf{k\neq0}}\hat{a}_{X,\mathbf{k}%
}e^{i\mathbf{k}.\mathbf{r}}\right]  $ with $\bar{A}=B$ and $\bar{B}=A$,
$f_{_{A}}=-e^{-i\varphi}\sin \left(  \frac{\theta}{2}\right)  $ and $f_{_{B}%
}=\cos \left(  \frac{\theta}{2}\right)  $. Following the Bogoliubov
approximation, we replace both $\hat{a}_{-}$ and $\hat{a}_{-}^{\dagger}$ by
$\sqrt{N}$, and substitute equation (\ref{eq4}) into $H-\mu N$ up to quadratic
order in $\tilde{a}_{\mathbf{r}}$. The terms linear in $\tilde{a}_{\mathbf{r}%
}$ or $\tilde{a}_{\mathbf{r}}^{\dagger}$ disappear due to the stability
condition of the condensate, and we arrive at the Bogoliubov Hamiltonian as
\begin{equation}
H-\mu N=\frac{1}{2}\mathcal{A}^{\dagger}M_{+}\mathcal{A}+\frac{1}{2}%
\sum_{\mathbf{k\neq0}}\hat{\alpha}_{\mathbf{k}}^{\dagger}M\left(
\mathbf{k}\right)  \hat{\alpha}_{\mathbf{k}} \label{eq0c}%
\end{equation}
with $\hat{\alpha}_{\mathbf{k}}^{\dagger}=\left(  \hat{a}_{A,\mathbf{k}%
}^{\dagger},\hat{a}_{B,\mathbf{k}}^{\dagger},\hat{a}_{A,-\mathbf{k}},\hat
{a}_{B,-\mathbf{k}}\right)  $ and $\mathcal{A}^{\dagger}=\left(  \hat{a}%
_{+}^{\dagger},\hat{a}_{+}\right)  $. Here, the $2\times2$ matrix $M_{+}$ and
$4\times4$ matrix $M\left(  \mathbf{k}\right)  $ are given by%
\begin{align}
M_{+}  &  =\left \{  h\left(  \mathbf{k}_{0}\right)  \left[  \cos \theta_{0}%
\cos \theta-\sin \theta_{0}\sin \theta \cos \left(  \varphi_{0}+\varphi \right)
\right]  -\mu \right. \nonumber \\
&  \left.  +8Un\left \vert f_{A}\right \vert ^{2}\left \vert f_{B}\right \vert
^{2}\right \}  I+N_{+},
\end{align}
with $N_{+}=4Un\left[  \operatorname{Re}\left(  f_{A}^{\ast}f_{B}^{\ast
}\right)  \sigma_{1}-\operatorname{Im}\left(  f_{A}^{\ast}f_{B}^{\ast}\right)
\sigma_{2}\right]  $, and%
\begin{align}
&  M\left(  \mathbf{k}\right) \nonumber \\
&  =\left(
\begin{array}
[c]{cc}%
\mathcal{H}\left(  \mathbf{k}\right)  -\mu I+4Un\left \vert F\right \vert ^{2} &
2UnF^{2}\\
2UnF^{\ast2} & \mathcal{H}^{T}\left(  -\mathbf{k}\right)  -\mu I+4Un\left \vert
F\right \vert ^{2}%
\end{array}
\right)  .
\end{align}
where $F=\mathrm{diag}\left(  f_{A},f_{B}\right)  $.

\begin{figure}[ptb]
\scalebox{0.48}{\includegraphics* [0.0in,0.0in][9.0in,5.6in]{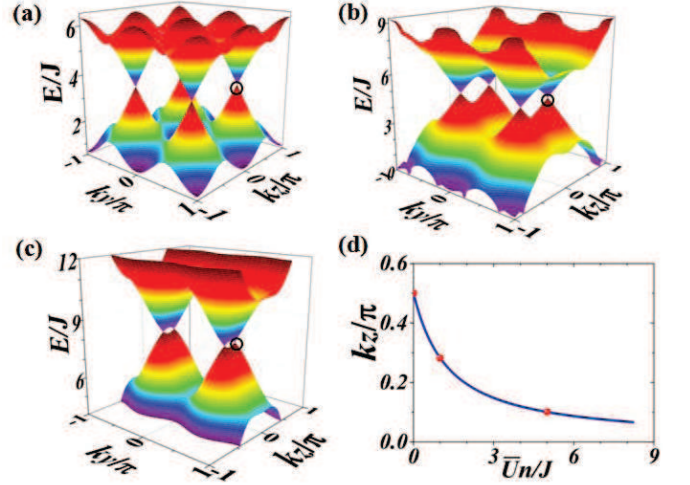}}\caption{(Color
online) The energy spectra of Bogoliubov quasiparticles with fixed $k_{x}=0$
and (a) $\overline{U}n/J=0.0$; (b) $\overline{U}n/J=1.0$; (c) $\overline
{U}n/J=5.0$. (d) The $k_{z}$-coordinates of Weyl points versus $\overline{U}%
n$, of which the ones indicated by black circles in (a), (b), (c) are
indicated by red points.}%
\label{dis}%
\end{figure}

To diagonalize above Bogoliubov Hamiltonian, we introduce paraunitary matrices
$W_{+}$ and $W_{\mathbf{k}}$ which satisfy $W_{+}^{\dagger}\sigma_{3}%
W_{+}=W_{+}\sigma_{3}W_{+}^{\dagger}=\sigma_{3}$, and $W_{\mathbf{k}}%
^{\dagger}\tau_{3}W_{\mathbf{k}}=W_{\mathbf{k}}\tau_{3}W_{\mathbf{k}}%
^{\dagger}=\tau_{3}$ with $\sigma_{3}=\mathrm{diag}\left(  1,-1\right)  $ and
$\tau_{3}=\mathrm{diag}\left(  1,1,-1,-1\right)  $ \cite{Fur,col}, and then
obtain%
\begin{equation}
W_{+}^{\dagger}M_{+}W_{+}=E_{+}\left(  0\right)  I,
\end{equation}%
\begin{equation}
W_{\mathbf{k}}^{\dagger}M\left(  \mathbf{k}\right)  W_{\mathbf{k}%
}=\mathrm{diag}\left(  E_{+,\mathbf{k}},E_{-,\mathbf{k}},E_{+,-\mathbf{k}%
},E_{-,-\mathbf{k}}\right)  .
\end{equation}
See Appendix \ref{Appendix_2} for details. At last, the Hamiltonian in Eq.
(\ref{eq0c}) is diagonalized as
\begin{equation}
H-\mu N=\sum_{\mathbf{k}}E_{+,\mathbf{k}}\hat{b}_{+,\mathbf{k}}^{\dagger}%
\hat{b}_{+,\mathbf{k}}+\sum_{\mathbf{k\neq0}}E_{-,\mathbf{k}}\hat
{b}_{-,\mathbf{k}}^{\dagger}\hat{b}_{-,\mathbf{k}}+const. \label{eq0d}%
\end{equation}
By direct numerical calculations, we get the Bogoliubov excitation bands
$E_{\pm,\mathbf{k}}$ as shown in Fig. \ref{dis}. It shows that there are Weyl
points in the excitation band. As the interaction strength increases, Weyl
points approach gradually with each other along the $k_{z}$-direction (see
Fig. \ref{dis} (d)).

To study the topological properties of excitation modes of Bogoliubov
quasiparticles, we define the basis vectors of $M\left(  \mathbf{k}\right)  $
as $\left \vert w_{\lambda}\left(  \mathbf{k}\right)  \right \rangle =\left(
\alpha_{A,\lambda}\left(  \mathbf{k}\right)  ,\alpha_{B,\lambda}\left(
\mathbf{k}\right)  ,\beta_{A,\lambda}\left(  \mathbf{k}\right)  ,\beta
_{B,\lambda}\left(  \mathbf{k}\right)  \right)  ^{T}$ with $\lambda=\pm$, of
which $\left \langle w_{\lambda^{\prime}}\left(  \mathbf{k}\right)  \right \vert
\tau_{3}\left \vert w_{\lambda}\left(  \mathbf{k}\right)  \right \rangle
=\delta_{\lambda^{\prime}\lambda}$. We then have $M\left(  \mathbf{k}\right)
\left \vert w_{\lambda}\left(  \mathbf{k}\right)  \right \rangle =E_{\lambda
}\left(  \mathbf{k}\right)  \tau_{3}\left \vert w_{\lambda}\left(
\mathbf{k}\right)  \right \rangle $, and the Berry curvature takes the form as
\begin{equation}
B_{\lambda,k}\left(  \mathbf{k}\right)  =i\epsilon_{ijk}\left \langle
\partial_{i}w_{\lambda}\left(  \mathbf{k}\right)  \left \vert \tau
_{3}\right \vert \partial_{j}w_{\lambda}\left(  \mathbf{k}\right)
\right \rangle
\end{equation}
with $\partial_{j}\equiv \partial/\partial k_{j}$ and $j=x$, $y$, $z$. The
topology of Weyl point is characterized by the first Chern number defined by
$C_{\mathbf{k}_{w},-}=%
%TCIMACRO{\doint }%
%BeginExpansion
{\displaystyle \oint}
%EndExpansion
d\mathbf{S\cdot B}_{-}\left(  \mathbf{k}\right)  $, which is calculated by the
integral of Berry curvature throughout the surface enclosing the Weyl point.
After direct calculations, we obtain $C_{\mathbf{k}_{w},-}=\pm1$ which implies
that Weyl points have different chiralities. It shows a synthetic magnetic
monopole located at $\left \{  \mathbf{k}_{w}\right \}  $.

\begin{figure}[ptb]
\scalebox{0.36}{\includegraphics* [0.0in,0.0in][11.3in,4.1in]{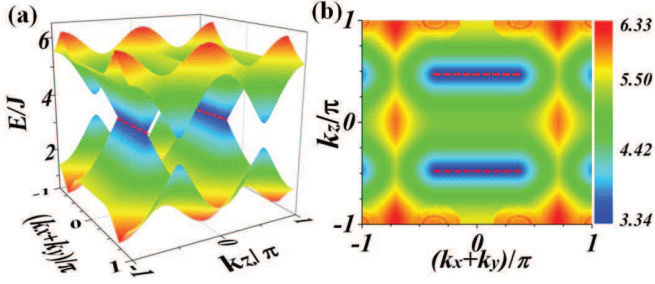}}\caption{(Color
online) (a) The Bogoliubov spectra of slab with finite width. The bosonic
surface-arc states are indicated by red dashed lines. (b) The contour plot of
upper energy spectra of excitation modes for BECs in a slab with finite width
along planes orthogonal to the $\vec{x}-\vec{y}$ direction. The arc states are
indicated by red dashed lines. In both (a) and (b), the parameter $\bar
{U}n/J=0.2$. }%
\label{edge1}%
\end{figure}

We apply the BdG theory to study the Bogoliubov excitations of superfluids by
choosing a slab with finite width along planes orthogonal to the $\vec{x}%
-\vec{y}$ direction, which has sharp the boundaries. The energy spectra and
contour plot of upper energy spectra of excitation modes are shown in Fig.
\ref{edge1}(a) and (b), respectively. There are topologically protected
surface states (dubbed bosonic arcs) of excitation modes which are the analogs
of Fermi arcs in electronic systems. As the interaction increases, two arcs
connected by two Weyl points will approach gradually with each other along the
$k_{z}$-direction.

\section{Excitations in Mott insulator phase}

\label{sec4}

In the strong coupling regime, we apply the path integral formulation to
calculate the excitation spectra of the MI state \cite{lim}. We first write
the partition function for the Bose-Hubbard extension in terms of path
integral as $Z=%
%TCIMACRO{\dint }%
%BeginExpansion
{\displaystyle \int}
%EndExpansion
\mathcal{D}a^{\ast}\mathcal{D}a\exp \left \{  -S\left[  a^{\ast},a\right]
/\hbar \right \}  $, where the action is given by%
\begin{align}
S\left[  a^{\ast},a\right]   &  =%
%TCIMACRO{\dint _{0}^{\beta}}%
%BeginExpansion
{\displaystyle \int_{0}^{\beta}}
%EndExpansion
d\tau \left[
%TCIMACRO{\dsum \limits_{\mathbf{r}}}%
%BeginExpansion
{\displaystyle \sum \limits_{\mathbf{r}}}
%EndExpansion
a_{\mathbf{r}}^{\ast}\left(  \tau \right)  (\hbar \partial_{\tau}-\mu
)a_{\mathbf{r}}\left(  \tau \right)  +H_{0}\left(  \tau \right)  \right.
\nonumber \\
&  \left.  +\frac{1}{2}U%
%TCIMACRO{\dsum \limits_{\mathbf{r}}}%
%BeginExpansion
{\displaystyle \sum \limits_{\mathbf{r}}}
%EndExpansion
a_{\mathbf{r}}^{\ast}\left(  \tau \right)  a_{\mathbf{r}}^{\ast}\left(
\tau \right)  a_{\mathbf{r}}\left(  \tau \right)  a_{\mathbf{r}}\left(
\tau \right)  \right]  \label{eq0e}%
\end{align}
with $\beta=\frac{1}{k_{B}T}$, $k_{B}$ is the Boltzman constant, and
$H_{0}\left(  \tau \right)  $ is obtained by replacing the bosonic operators
$\left(  \hat{a}_{\mathbf{r}}^{\dagger}\text{, }\hat{a}_{\mathbf{r}}\right)  $
in Eq. (\ref{eq0}) by complex functions $\left(  a_{\mathbf{r}}^{\ast}\left(
\tau \right)  \text{, }a_{\mathbf{r}}\left(  \tau \right)  \right)  $. To
decouple the hopping terms, we make use of Hubbard-Stratonovich transformation
and rewrite the action as \cite{lim}%

\begin{equation}
S\left[  a^{\ast},a,\psi^{\ast},\psi \right]  =S\left[  a^{\ast},a\right]
+\int_{0}^{\hbar \beta}d\tau%
%TCIMACRO{\dsum \limits_{\mathbf{r,j}}}%
%BeginExpansion
{\displaystyle \sum \limits_{\mathbf{r,j}}}
%EndExpansion
(\psi_{\mathbf{r}}^{\ast}-a_{\mathbf{r}}^{\ast})J_{\mathbf{rj}}(\psi
_{\mathbf{j}}-a_{\mathbf{j}}), \label{eq0f}%
\end{equation}
where $\psi_{\mathbf{r}}^{\ast}$ and $\psi_{\mathbf{j}}$ are the order
parameter fields, and $J_{\mathbf{rj}}$ denotes real nearest-neighbor hopping
parameters along the $x$, $y$, and $z$ direction. By performing integration
over the complex fields $a_{\mathbf{r}}^{\ast}$ and $a_{\mathbf{r}}$, and
after direct calculations, we get the effective action up to the second order
near the phase transition point as%

\begin{align}
S^{eff}\left[  \psi^{\ast},\psi \right]   &  =%
%TCIMACRO{\dsum \limits_{\mathbf{k},\omega_{m}}}%
%BeginExpansion
{\displaystyle \sum \limits_{\mathbf{k},\omega_{m}}}
%EndExpansion
\Psi_{\mathbf{k},\omega_{m}}^{\ast}(\mathcal{H}\left(  \mathbf{k}\right)
\mathcal{-H}\left(  \mathbf{k}\right)  ^{2}f_{\omega_{m}})\Psi_{\mathbf{k}%
,\omega_{m}}\nonumber \\
&  =%
%TCIMACRO{\dsum \limits_{\mathbf{k},\omega_{m}}}%
%BeginExpansion
{\displaystyle \sum \limits_{\mathbf{k},\omega_{m}}}
%EndExpansion
\Psi_{\mathbf{k},\omega_{m}}^{\ast}\left[  -\hbar G^{-1}(\mathbf{k}%
,i\omega_{m})\right]  \Psi_{\mathbf{k},\omega_{m}}, \label{eq0g}%
\end{align}
where $\Psi_{\mathbf{k},\omega_{m}}^{\ast}=\left(  \psi_{A\mathbf{k}%
,\omega_{m}}^{\ast}\text{, }\psi_{B\mathbf{k},\omega_{m}}^{\ast}\right)  $, and%

\begin{equation}
f_{\omega_{m}}=\frac{\tilde{n}+1}{-i\hbar \omega_{m}-\mu+\tilde{n}U}%
+\frac{\tilde{n}}{i\hbar \omega_{m}+\mu-(\tilde{n}-1)U}.
\end{equation}
See Appendix \ref{Appendix_3} for detailed calculations.

\begin{figure}[ptb]
\scalebox{0.46}{\includegraphics* [0.0in,0.0in][10.0in,5.6in]{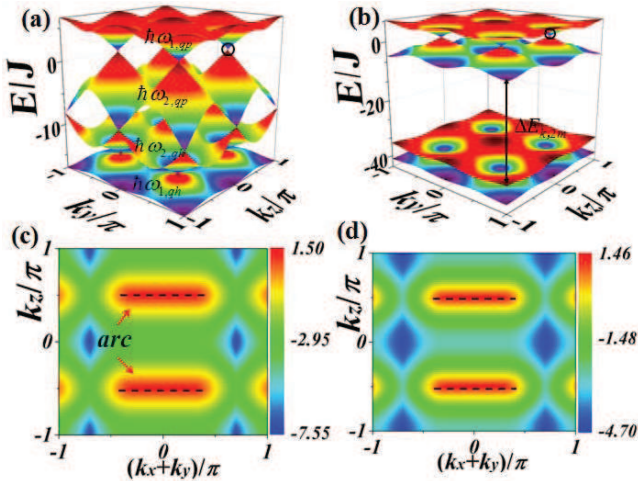}}\caption{(Color
online) The energy spectra of quasiparticle and quasihole excitations with
fixed $k_{x}=0$ for (a) $\bar{U}/J=8.25$ and (b) $\bar{U}/J=20.0$ in MI$(1,1)$
phase; and the contour plot of energy spectra of lower one of
quasiparticle-excitation modes in MI$(1,1)$ phase in a slab with finite width
along planes orthogonal to the $\vec{x}-\vec{y}$ direction for (c) $\bar
{U}/J=8.25$ and (d) $\bar{U}/J=20.0$ in MI$(1,1)$ phase. The locations of the
Weyl nodes are shown indicated by black circles in (a) and (b). The bosonic
arc states are indicated by black dashed lines in (c) and (d). }%
\label{marc}%
\end{figure}

Under the usual analytic continuation $i\omega_{m}\rightarrow \omega_{m}$, we
can obtain a function of real energies $\hbar \omega$, i.e.,%

\begin{equation}
\det \left[  G^{-1}\left(  \mathbf{k},i\omega_{m}\right)  \right]  =0.
\end{equation}
The quasiparticle- and quasihole-dispersion relations are obtained as
\begin{align}
\hbar \omega_{1,qp,ph} &  =\frac{1}{2}\left[  -2\mu+(2\tilde{n}-1)U-\frac
{1}{f_{\omega_{m_{+}}}}\pm \Delta E_{\mathbf{k}1}\right]  ,\nonumber \\
\hbar \omega_{2,qp,ph} &  =\frac{1}{2}\left[  -2\mu+(2\tilde{n}-1)U-\frac
{1}{f_{\omega_{m_{-}}}}\pm \Delta E_{\mathbf{k}2}\right]  ,\label{disp}%
\end{align}
where $\Delta E_{\mathbf{k},1}=\sqrt{U^{2}-\frac{(4\tilde{n}+2)U}%
{f_{\omega_{m_{+}}}}+\frac{1}{f_{\omega_{m_{+}}}^{2}}}$, $\Delta
E_{\mathbf{k},2}=\sqrt{U^{2}-\frac{(4\tilde{n}+2)U}{f_{\omega_{m_{-}}}}%
+\frac{1}{f_{\omega_{m_{-}}}^{2}}}$, $f_{\omega_{m_{\pm}}}=\frac{A+B\pm
\sqrt{(A-B)^{2}+4C^{\ast}C}}{2(AB-C^{\ast}C)}$, $A=-B=2J_{z}\cos(k_{z})$ and
$C=-2J_{y}\cos(k_{y})+i2J_{x}\sin(k_{x})$. See quasiparticle- and
quasihole-dispersions in Fig. \ref{marc} (a) and (b). In addition, we also
plot the first-order approximations (the minimum values of $\Delta
E_{\mathbf{k},1}$ and $\Delta E_{\mathbf{k},2}$, i.e., $\Delta E_{\mathbf{k}%
,1m}$ and $\Delta E_{\mathbf{k},2m}$) to the dispersion of the density
fluctuations in Fig. \ref{mi1} (b). Fig. \ref{mi1} (b) indicates that the band
gap disappears as we approach the critical value $\bar{U}_{c}/J\approx8.25$
that is consistent with the result found in Fig. \ref{mi1} (a). Fig.
\ref{marc} (a) and (b) also show that there are Weyl nodes in the Bogoliubov
excitation spectra in Mott-insulator phase, and the Mott gap $E_{\mathbf{k}%
,2m}$ becomes larger as the interaction strength increases similar to that in
conventional Bose-Hubbard models.

Next, we apply the BdG theory to study the quasiparticle and quasihole
excitations in Mott-insulator phase by choosing a slab with finite width and
the sharp boundaries along planes orthogonal to the $\vec{x}-\vec{y}$
direction. After calculations, we present the results in Fig. \ref{marc} (c)
and (d), which show that there are also bosonic surface-arc states at boundaries.

\section{Discussion and Conclusion}

\label{sec5}

The Hamiltonian of WSM in Eq. (\ref{eq0}) in three-dimensional optical
lattices has been proposed by using laser-assisted tunneling in Ref.
\cite{ten}. The interaction between particles may be tuned readily by using
Feshbach resonance technique. To experimentally measure the topological
properties of the elementary Bogoliubov excitations in SF phase, one may
coherently transfer a small portion of the condensate into a surface mode by
stimulated Raman transitions \cite{ern,zhi}. Owing to that the original bosons
and the Bogoliubov excitations is connected by the paraunitary matrix, a
density wave may form from the interference of the surface modes and the
condensate wave function by the mechanism discussed in Ref. \cite{Fur}.

In summary, Bogoliubov excitations in Bose-Hubbard extension of the WSM are
studied. By using Bogoliubov theory, we calculate the energy spectra of
excitation modes for the system with weak repulsive interactions, and find
their non-trivial properties owing to the existence of Weyl points. There
exist bosonic surface-arcs connected by Weyl points with different chiralities
analogs of Fermi arc in WSM of electronic system. As the interaction
increases, Weyl points approach gradually with each other along the $k_{z}%
$-direction. In the strong coupling regime, the system is in MI phase. By
using path integral formulation, we find that there are two quasiparticle
dispersions touching at stable nodes (Weyl points), and there are also bosonic
surface-arc at boundaries.

In addition to the type-I WSM considered in this paper, we expect that there
are also novel excitations in the Bose-Hubbard extension of type-II
\cite{xu1,sol,xu2} and hybrid WSMs \cite{li,kong}. The bosonic Weyl
excitations will deepen our standing of quantum many body physics in boson systems.

\begin{acknowledgments}
This work is supported by NSFC under the grant No. 11504285, 11474025,
11674026, SRFDP, the Scientific Research Program Funded by Shaanxi Provincial
Education Department under the grant No. 15JK1348, and supported by Young
Talent fund of University Association for Science and Technology in Shaanxi, China.
\end{acknowledgments}

\appendix

\section{Landau theory for superfluid-Mott insulator transition}

\label{Appendix_1}

In the strong coupling limit, we first introduce a local superfluid order
parameter given by \cite{van,lim}
\begin{equation}
\psi_{\mathbf{r}}=\langle \hat{a}_{\mathbf{r}}^{\dag}\rangle=\langle \hat
{a}_{\mathbf{r}}\rangle.
\end{equation}
Due to $A$-sublattice and $B$-sublattice for the lattice system, the order
parameters are defined as $\psi_{\mathbf{r\in}A}\equiv \psi_{A}$,
$\psi_{\mathbf{r\in}B}\equiv \psi_{B}$, $n_{\mathbf{r\in}A}\equiv n_{A}$ and
$n_{\mathbf{r\in}B}\equiv n_{B}$, respectively. With the help of
$\psi_{\mathbf{r}}$, we decouple the hopping term into%

\begin{equation}
\hat{a}_{\mathbf{r}}^{\dag}\hat{a}_{\mathbf{j}}=\hat{a}_{\mathbf{r}}^{\dag
}\psi_{\mathbf{j}}+\psi_{\mathbf{r}}\hat{a}_{\mathbf{j}}-\psi_{\mathbf{r}}%
\psi_{\mathbf{j}},
\end{equation}
where $\mathbf{j}$ denotes the coordinate of lattice site $\mathbf{r}$'s
nearest neighbor site. Then the Hamiltonian takes following form:%

\begin{align}
\hat{H}^{eff}  &  =\sum_{\mathbf{r}\in B}\left[  -J_{x}(\hat{a}_{\mathbf{r}%
}+\hat{a}_{\mathbf{r}}^{\dag}-\psi_{\mathbf{r}})\psi_{\mathbf{r}%
+\mathbf{\delta}_{x}}\right. \nonumber \\
&  \left.  -J_{y}(\hat{a}_{\mathbf{r}}+\hat{a}_{\mathbf{r}}^{\dag}%
-\psi_{\mathbf{r}})\psi_{\mathbf{r}+\mathbf{\delta}_{y}}-J_{z}(\hat
{a}_{\mathbf{r}}+\hat{a}_{\mathbf{r}}^{\dag}-\psi_{\mathbf{r}})\psi
_{\mathbf{r}+\mathbf{\delta}_{z}})\right] \nonumber \\
&  +\sum_{\mathbf{r}\in A}\left[  J_{x}(\hat{a}_{\mathbf{r}}+\hat
{a}_{\mathbf{r}}^{\dag}-\psi_{\mathbf{r}})\psi_{\mathbf{r}+\mathbf{\delta}%
_{x}}\right. \nonumber \\
&  \left.  -J_{y}(\hat{a}_{\mathbf{r}}+\hat{a}_{\mathbf{r}}^{\dag}%
-\psi_{\mathbf{r}})\psi_{\mathbf{r}+\mathbf{\delta}_{y}}+J_{z}(\hat
{a}_{\mathbf{r}}+\hat{a}_{\mathbf{r}}^{\dag}-\psi_{\mathbf{r}})\psi
_{\mathbf{r}+\mathbf{\delta}_{z}})\right] \nonumber \\
&  +\frac{U}{2}\sum_{\mathbf{r}}n_{\mathbf{r}}(n_{\mathbf{r}}-1)-\mu
\sum_{\mathbf{r}}\hat{a}_{\mathbf{r}}^{\dag}\hat{a}_{\mathbf{r}}.
\end{align}
For the system, the effective onsite Hamiltonian $\hat{H}_{on}^{eff}$ is then
given by%

\begin{align}
\hat{H}_{on}^{eff}  &  =J_{x}(\hat{a}_{A}^{\dag}+\hat{a}_{A})\psi_{B}%
-J_{x}(\hat{a}_{B}^{\dag}+\hat{a}_{B})\psi_{1}\nonumber \\
&  -J_{y}(\hat{a}_{A}^{\dag}+\hat{a}_{A})\psi_{B}-J_{y}(\hat{a}_{B}^{\dag
}+\hat{a}_{B})\psi_{A}+2J_{y}\psi_{A}\psi_{B}\nonumber \\
&  +J_{z}(\hat{a}_{A}^{\dag}+\hat{a}_{A})\psi_{A}-J_{z}(\hat{a}_{B}^{\dag
}+\hat{a}_{B})\psi_{B}\nonumber \\
&  +J_{z}(\psi_{B}^{2}-\psi_{A}^{2})+\frac{\overline{U}}{2}(n_{A}^{2}%
+n_{B}^{2}-n_{u})-\overline{\mu}n_{u}%
\end{align}
where $\bar{U}=U/z$, and $\bar{\mu}=\bar{\mu}/z$, and $z=2$ is the number of
nearest-neighbor sites in one direction. Next, we write $\hat{H}_{on}=\hat
{H}_{on}^{0}+\hat{H}_{on}^{^{\prime}}$ with
\begin{align}
\hat{H}_{on}^{0}  &  =\frac{\overline{U}}{2}(n_{A}^{2}+n_{B}^{2}%
-n_{u})-\overline{\mu}n_{u}\nonumber \\
&  +2J_{y}\psi_{A}\psi_{B}+J_{z}(\psi_{B}^{2}-\psi_{A}^{2})
\end{align}
and%

\begin{align}
\hat{H}_{on}^{\prime}  &  =J_{x}(\hat{a}_{A}^{\dag}+\hat{a}_{A})\psi_{B}%
-J_{x}(\hat{a}_{B}^{\dag}+\hat{a}_{B})\psi_{A}\nonumber \\
&  -J_{y}(\hat{a}_{A}^{\dag}+\hat{a}_{A})\psi_{B}-J_{y}(\hat{a}_{B}^{\dag
}+\hat{a}_{B})\psi_{A}\nonumber \\
&  +J_{z}(\hat{a}_{A}^{\dag}+\hat{a}_{A})\psi_{A}-J_{z}(\hat{a}_{B}^{\dag
}+\hat{a}_{B})\psi_{B}.
\end{align}
In the occupation numbers basis, we can find that the odd powers of the
expansion of energy are always zero. Hence, the energy for the zero-order
terms is then given by%

\begin{align}
E_{g}^{(0)}  &  =\min(e_{\{n_{A};n_{B}\}}^{(0)})\nonumber \\
&  =\frac{\overline{U}}{2}(n_{A}^{2}+n_{B}^{2}-n_{u})-\overline{\mu}%
n_{u}\nonumber \\
&  +2J_{y}\psi_{A}\psi_{B}+J_{z}(\psi_{B}^{2}-\psi_{A}^{2}),
\end{align}
and the energy for the second-order perturbation is%

\begin{equation}
E_{g}^{(2)}=\frac{\left \langle n_{A};n_{B}\left \vert H_{on}^{^{\prime}%
}\right \vert k_{1};k_{2}\right \rangle \left \langle k_{1};k_{2}\left \vert
H_{on}^{^{\prime}}\right \vert n_{A};n_{B}\right \rangle }{E_{g}^{(0)}%
-E_{k}^{(0)}},
\end{equation}
where $k_{1}=n_{A}+1$, $k_{2}=n_{B}$ or $k_{1}=n_{A}$, $k_{2}=n_{B}+1$. After
direct calculations, we obtain%

\begin{align}
E_{g}^{(2)}  &  =\frac{n_{A}(J_{x}\psi_{B}-J_{y}\psi_{B}+J_{z}\psi_{A})^{2}%
}{(n_{A}-1)\overline{U}-\overline{\mu}}\nonumber \\
&  +\frac{(n_{A}+1)(J_{x}\psi_{B}-J_{y}\psi_{B}+J_{z}\psi_{A})^{2}}%
{-n_{A}\overline{U}+\overline{\mu}}\nonumber \\
&  +\frac{n_{B}(-J_{x}\psi_{A}-J_{y}\psi_{A}-J_{z}\psi_{B})^{2}}%
{(n_{B}-1)\overline{U}-\overline{\mu}}\nonumber \\
&  +\frac{(n_{B}+1)(-J_{x}\psi_{A}-J_{y}\psi_{A}-J_{z}\psi_{B})^{2}}%
{-n_{B}\overline{U}+\overline{\mu}}%
\end{align}

In summary, the per unit cell ground-state energy for the system in terms of
$\left(  \psi_{A},\psi_{B}\right)  $ is given by%

\begin{align}
E_{g}  &  \simeq E_{g}^{(0)}+E_{g}^{(2)}+...\nonumber \\
&  =a_{0}+a_{2}\psi_{A}^{2}+c_{2}\psi_{A}\psi_{B}+b_{2}\psi_{B}^{2}%
+\mathcal{O}\left(  \psi_{A}^{4},\psi_{B}^{4}\right)  ,
\end{align}
where the coefficients $a_{0}$, $a_{2}$, $c_{2}$, and $b_{2}$ are listed in
Eqs. (\ref{eqq1})-(\ref{eqq4}).

\section{Diagonalization of Bogoliubov Hamiltonian for bosons}

\label{Appendix_2}

To diagonalize Bogoliubov Hamiltonian in Eq. (\ref{eq0c}), we perform
generalized Bogoliubov transformations as%
\begin{equation}
\left(
\begin{array}
[c]{c}%
\hat{a}_{+}\\
\hat{a}_{+}^{\dagger}%
\end{array}
\right)  =W_{+}\left(
\begin{array}
[c]{c}%
\hat{b}_{+}\left(  0\right) \\
\hat{b}_{+}^{\dagger}\left(  0\right)
\end{array}
\right)  ,\hat{\alpha}_{\mathbf{k}}=W\left(  \mathbf{k}\right)  \hat{\beta
}_{\mathbf{k}}%
\end{equation}
with
\[
\hat{\beta}_{\mathbf{k}}^{\dagger}=\left(  \hat{b}_{+,\mathbf{k}}^{\dagger
},\hat{b}_{-,\mathbf{k}}^{\dagger},\hat{b}_{+,-\mathbf{k}},\hat{b}%
_{-,-\mathbf{k}}\right)  .
\]
Here, $W_{+}$ and $W_{\mathbf{k}}$ are paraunitary matrices that satisfy
following conditions:%
\begin{align}
W_{+}^{\dagger}\sigma_{3}W_{+}  &  =W_{+}\sigma_{3}W_{+}^{\dagger}=\sigma
_{3},\nonumber \\
W_{\mathbf{k}}^{\dagger}\tau_{3}W_{\mathbf{k}}  &  =W_{\mathbf{k}}\tau
_{3}W_{\mathbf{k}}^{\dagger}=\tau_{3},
\end{align}
where $\sigma_{3}=\mathrm{diag}\left(  1,-1\right)  $ and $\tau_{3}%
=\mathrm{diag}\left(  1,1,-1,-1\right)  $. The $4\times4$ paraunitary matrix
$W_{\mathbf{k}}$ can be constructed numerically, which is written as%
\begin{equation}
W_{\mathbf{k}}=\left(
\begin{array}
[c]{cc}%
U\left(  \mathbf{k}\right)  & V^{\ast}\left(  -\mathbf{k}\right) \\
V\left(  \mathbf{k}\right)  & U^{\ast}\left(  -\mathbf{k}\right)
\end{array}
\right)  ,
\end{equation}
where
\begin{equation}
U\left(  \mathbf{k}\right)  =\left(
\begin{array}
[c]{cc}%
\alpha_{A,+}\left(  \mathbf{k}\right)  & \alpha_{A,-}\left(  \mathbf{k}\right)
\\
\alpha_{B,+}\left(  \mathbf{k}\right)  & \alpha_{B,-}\left(  \mathbf{k}%
\right)
\end{array}
\right)  ,
\end{equation}
and
\begin{equation}
V\left(  \mathbf{k}\right)  =\left(
\begin{array}
[c]{cc}%
\beta_{A,+}\left(  \mathbf{k}\right)  & \beta_{A,-}\left(  \mathbf{k}\right)
\\
\beta_{B,+}\left(  \mathbf{k}\right)  & \beta_{B,-}\left(  \mathbf{k}\right)
\end{array}
\right)  .
\end{equation}
The transformations with $W_{+}$ and $W_{\mathbf{k}}$ ensure the invariance of
the bosonic commutation relations for Bogoliubov excitations. With the help of
them, we may diagonalize the Bogoliubov Hamiltonian in Eq. (\ref{eq0c})
readily, and at last arrive at Eq. (\ref{eq0d}).

\section{Collective modes in MI phase}

\label{Appendix_3}

The partition function for the Bose-Hubbard extension in terms of path
integral is written as $Z=%
%TCIMACRO{\dint }%
%BeginExpansion
{\displaystyle \int}
%EndExpansion
\mathcal{D}a^{\ast}\mathcal{D}a\exp \left \{  -S\left[  a^{\ast},a\right]
/\hbar \right \}  $, where the action is shown in Eq. (\ref{eq0e}). For
convenience, in the following, we denote $H_{0}\left(  \tau \right)  $ (hopping
terms) as $\sum_{\mathbf{r,j}}J_{\mathbf{rj}}a_{\mathbf{r}}^{\ast
}a_{\mathbf{j}}$ that are perturbation terms in strong coupling regime, where
$J_{\mathbf{rj}}$ denotes real nearest-neighbor hopping parameters along the
$x$, $y$, and $z$ direction. By using Hubbard-Stratonovich transformation, we
obtain the action as shown in Eq. (\ref{eq0f}). After direct calculations, the
action becomes%

\begin{align}
S\left[  a^{\ast},a,\psi^{\ast},\psi \right]   &  =\int_{0}^{\hbar \beta}%
d\tau \left[
%TCIMACRO{\dsum \limits_{\mathbf{r}}}%
%BeginExpansion
{\displaystyle \sum \limits_{\mathbf{r}}}
%EndExpansion
a_{\mathbf{r}}^{\ast}(\hbar \frac{\partial}{\partial \tau}-\mu)a_{\mathbf{r}%
}+\frac{1}{2}U%
%TCIMACRO{\dsum \limits_{\mathbf{r}}}%
%BeginExpansion
{\displaystyle \sum \limits_{\mathbf{r}}}
%EndExpansion
a_{\mathbf{r}}^{\ast2}\right. \nonumber \\
&  \left.  \times a_{\mathbf{r}}^{2}-%
%TCIMACRO{\dsum \limits_{\mathbf{r,j}}}%
%BeginExpansion
{\displaystyle \sum \limits_{\mathbf{r,j}}}
%EndExpansion
J_{\mathbf{rj}}(a_{\mathbf{r}}^{\ast}\psi_{\mathbf{j}}+\psi_{\mathbf{r}}%
^{\ast}a_{\mathbf{j}})+%
%TCIMACRO{\dsum \limits_{\mathbf{r,j}}}%
%BeginExpansion
{\displaystyle \sum \limits_{\mathbf{r,j}}}
%EndExpansion
J_{\mathbf{rj}}\psi_{\mathbf{r}}^{\ast}\psi_{\mathbf{j}}\right]  .
\end{align}
Then, we have the explicit form as follows: \begin{widetext}%
\begin{align}
e^{-S^{eff\left[  \psi^{\ast},\psi \right]  }} &  =\exp \left(  -\frac{1}{\hbar
}\int_{0}^{\hbar \beta}d\tau%
%TCIMACRO{\dsum \limits_{\mathbf{r,j}}}%
%BeginExpansion
{\displaystyle \sum \limits_{\mathbf{r,j}}}
%EndExpansion
J_{\mathbf{rj}}\psi_{\mathbf{r}}^{\ast}(\tau)\psi_{\mathbf{j}}(\tau)\right)
\nonumber \\
\times &
%TCIMACRO{\dint }%
%BeginExpansion
{\displaystyle \int}
%EndExpansion
\mathcal{D}a^{\ast}\mathcal{D}a\exp \left \{  -S^{(0)}\left[  a^{\ast},a\right]
/\hbar-\frac{1}{\hbar}\int_{0}^{\hbar \beta}d\tau \left(  -%
%TCIMACRO{\dsum \limits_{\mathbf{r,j}}}%
%BeginExpansion
{\displaystyle \sum \limits_{\mathbf{r,j}}}
%EndExpansion
J_{\mathbf{rj}}(a_{\mathbf{r}}^{\ast}(\tau)\psi_{\mathbf{j}}(\tau
)+\psi_{\mathbf{r}}^{\ast}(\tau)a_{\mathbf{j}}(\tau))\right)  \right \}  ,
\end{align}
\end{widetext}where we have denoted the action for $J_{\mathbf{rj}}=0$ by
$S^{(0)}\left[  a^{\ast},a\right]  $.

Now by using the relation $\left \langle e^{A_{i}}\right \rangle
=e^{\left \langle A_{i}\right \rangle +\frac{1}{2}(\left \langle A_{i}%
^{2}\right \rangle -\left \langle A_{i}\right \rangle ^{2})+...}$, we can get the
expression for the action $S^{eff}\left[  \psi^{\ast},\psi \right]  $ up to the
second order, i.e.,%

\[
S^{eff}\left[  \psi^{\ast},\psi \right]  \approx S^{(0)}\left[  \psi^{\ast
},\psi \right]  +S^{(2)}\left[  \psi^{\ast},\psi \right]  ,
\]
where $S^{(2)}\left[  \psi^{\ast},\psi \right]  $ is given by
\begin{widetext}%
\begin{align}
S^{(2)}\left[  \psi^{\ast},\psi \right]   &  =\int_{0}^{\hbar \beta}d\tau%
%TCIMACRO{\dsum \limits_{\mathbf{r,j}}}%
%BeginExpansion
{\displaystyle \sum \limits_{\mathbf{r,j}}}
%EndExpansion
J_{\mathbf{rj}}\psi_{\mathbf{r}}^{\ast}(\tau)\psi_{\mathbf{j}}(\tau)-\frac
{1}{2\hbar}\left \langle \left(  \int_{0}^{\hbar \beta}d\tau%
%TCIMACRO{\dsum \limits_{\mathbf{r,j}}}%
%BeginExpansion
{\displaystyle \sum \limits_{\mathbf{r,j}}}
%EndExpansion
J_{\mathbf{rj}}\left[  a_{\mathbf{r}}^{\ast}(\tau)\psi_{\mathbf{j}}(\tau
)+\psi_{\mathbf{r}}^{\ast}(\tau)a_{\mathbf{j}}(\tau)\right]  \right)
^{2}\right \rangle _{S^{(0)}}\nonumber \\
&  =-\frac{1}{2\hbar}\left \langle \int_{0}^{\hbar \beta}\int_{0}^{\hbar \beta
}d\tau d\tau^{^{\prime}}%
%TCIMACRO{\dsum \limits_{\mathbf{r,j},\mathbf{r}^{\prime}\mathbf{,j}^{\prime}%
%}}%
%BeginExpansion
{\displaystyle \sum \limits_{\mathbf{r,j},\mathbf{r}^{\prime}\mathbf{,j}%
^{\prime}}}
%EndExpansion
J_{\mathbf{rj}}J_{\mathbf{r}^{\prime}\mathbf{j}^{\prime}}\left[
a_{\mathbf{r}}^{\ast}(\tau)\psi_{\mathbf{j}}(\tau)+\psi_{\mathbf{r}}^{\ast
}(\tau)a_{\mathbf{j}}(\tau)\right]  \left[  a_{\mathbf{r}^{^{\prime}}}^{\ast
}(\tau^{^{\prime}})\psi_{\mathbf{j}^{^{\prime}}}(\tau^{^{\prime}}%
)+\psi_{\mathbf{r}^{^{\prime}}}^{\ast}(\tau^{^{\prime}})b_{\mathbf{j}%
^{^{\prime}}}(\tau^{^{\prime}})\right]  \right \rangle _{S^{(0)}}\nonumber \\
&  +\int_{0}^{\hbar \beta}d\tau%
%TCIMACRO{\dsum \limits_{\mathbf{r,j}}}%
%BeginExpansion
{\displaystyle \sum \limits_{\mathbf{r,j}}}
%EndExpansion
J_{\mathbf{rj}}\psi_{\mathbf{r}}^{\ast}(\tau)\psi_{\mathbf{j}}(\tau).
\end{align}
\end{widetext}According to the correlations, i.e., $\left \langle
a_{\mathbf{r}}^{\ast}a_{\mathbf{j}}^{\ast}\right \rangle _{S^{(0)}%
}=\left \langle a_{\mathbf{r}}a_{\mathbf{j}}\right \rangle _{S^{(0)}}=0$ and
$\left \langle a_{\mathbf{r}}^{\ast}a_{\mathbf{j}}\right \rangle _{S^{(0)}%
}=\left \langle a_{\mathbf{r}}a_{\mathbf{j}}^{\ast}\right \rangle _{S^{(0)}%
}=\left \langle a_{\mathbf{r}}a_{\mathbf{j}}^{\ast}\right \rangle _{S^{(0)}%
}\delta_{\mathbf{rj}}$, we have the action $S^{(2)}\left[  \psi^{\ast}%
,\psi \right]  $, i.e., \begin{widetext}
\begin{equation}
S^{(2)}\left[  \psi^{\ast},\psi \right]  =\int_{0}^{\hbar \beta}d\tau%
%TCIMACRO{\dsum \limits_{\mathbf{rj}}}%
%BeginExpansion
{\displaystyle \sum \limits_{\mathbf{rj}}}
%EndExpansion
\psi_{\mathbf{r}}^{\ast}J_{\mathbf{rj}}\psi_{\mathbf{j}}-\frac{1}{\hbar}%
\int_{0}^{\hbar \beta}\int_{0}^{\hbar \beta}d\tau d\tau^{^{\prime}}%
%TCIMACRO{\dsum \limits_{\mathbf{r,j},\mathbf{r}^{\prime}\mathbf{,j}^{\prime}%
%}}%
%BeginExpansion
{\displaystyle \sum \limits_{\mathbf{r,j},\mathbf{r}^{\prime}\mathbf{,j}%
^{\prime}}}
%EndExpansion
J_{\mathbf{rj}}J_{\mathbf{r}^{\prime}\mathbf{j}^{\prime}}\psi_{\mathbf{j}%
}^{\ast}(\tau)\left \langle T_{\tau}\left[  a_{\mathbf{r}}(\tau)a_{\mathbf{r}%
^{^{\prime}}}^{\ast}(\tau^{^{\prime}})\right]  \right \rangle _{S^{(0)}}%
\psi_{\mathbf{j}^{^{\prime}}}(\tau^{^{\prime}}).
\end{equation}
\end{widetext}For the quadratic term, we can get the formulation in momentum
space by using Flourier transformation, i.e.,%

\begin{equation}%
%TCIMACRO{\dsum \limits_{ijj^{^{\prime}}}}%
%BeginExpansion
{\displaystyle \sum \limits_{ijj^{^{\prime}}}}
%EndExpansion
t_{ij}t_{ij^{^{\prime}}}\psi_{j}^{\ast}(\tau)\psi_{j^{^{\prime}}}%
(\tau^{^{\prime}})=%
%TCIMACRO{\dsum \limits_{\mathbf{k}}}%
%BeginExpansion
{\displaystyle \sum \limits_{\mathbf{k}}}
%EndExpansion
\Psi_{\mathbf{k}}^{\ast}\left(  \tau \right)  \mathcal{H}\left(  \mathbf{k}%
\right)  ^{2}\Psi_{\mathbf{k}}^{\ast}\left(  \tau^{\prime}\right)  ,
\end{equation}
where $\Psi_{\mathbf{k}}^{\ast}\left(  \tau \right)  =(\psi_{A\mathbf{k}}%
^{\ast}(\tau),\psi_{B\mathbf{k}}^{\ast}(\tau))$, and the Hamiltonian matrix in
Eq. (\ref{eq2}) is re-written as
\begin{equation}
\mathcal{H}\left(  \mathbf{k}\right)  =\left(
\begin{array}
[c]{cc}%
A & C^{\ast}\\
C & B
\end{array}
\right)
\end{equation}
with $A=-B=2J_{z}\cos(k_{z})$ and $C=-2J_{y}\cos(k_{y})+i2J_{x}\sin(k_{x})$.
Near the phase transformation point, the zero-order effective action $S^{(0)}$
is going to zero. Therefore, the effective action becomes
%\begin{widetext}%
\begin{align}
S^{eff} &  =\int_{0}^{\hbar \beta}%
%TCIMACRO{\dsum \limits_{k}}%
%BeginExpansion
{\displaystyle \sum \limits_{k}}
%EndExpansion
\Psi_{\mathbf{k}}^{\ast}\left(  \tau \right)  \mathcal{H}\Psi_{\mathbf{k}%
}\left(  \tau^{\prime}\right)  \nonumber \\
&  -\frac{1}{\hbar}\int_{0}^{\hbar \beta}\int_{0}^{\hbar \beta}d\tau
d\tau^{^{\prime}}\left \langle a_{\mathbf{r}}(\tau)a_{\mathbf{r}^{^{\prime}}%
}^{\ast}(\tau^{^{\prime}})\right \rangle
%TCIMACRO{\dsum \limits_{\mathbf{k}}}%
%BeginExpansion
{\displaystyle \sum \limits_{\mathbf{k}}}
%EndExpansion
\Psi_{\mathbf{k}}^{\ast}\mathcal{H}^{2}\Psi_{\mathbf{k}}%
\end{align}
%\end{widetext}
with $\left \langle a_{\mathbf{r}}(\tau)a_{\mathbf{r}^{^{\prime}}}^{\ast}%
(\tau^{^{\prime}})\right \rangle _{S^{(0)}}=\left \langle T_{\tau}\left[
a_{\mathbf{r}}(\tau)a_{\mathbf{r}^{^{\prime}}}^{\ast}(\tau^{^{\prime}%
})\right]  \right \rangle _{S^{(0)}}$. Because the time ordering can be
expressed by Matsubara Green function, i.e.,

%\begin{figure*}[ptb]
%\scalebox{0.65}{\includegraphics* [0.1in,0.0in][10.5in,5.3in]{fig7.eps}}\caption{(Color
%online) The contour plot of energy spectra of lower one of
%quasiparticle-excitation modes in MI$(1,1)$ phase in a slab with finite width
%along planes orthogonal to the $\vec{x}-\vec{y}$ direction. The parameters
%$\bar{U}/J$ are chosen as (a) 8.25; (b)8.5; (c) 8.75; (d) 9.0; (e) 10.0;
%(f) 20.0.}%
%\label{marcc}%
%\end{figure*}%
%

\begin{align}
\left \langle T_{\tau}\left[  a_{\mathbf{r}}(\tau)a_{\mathbf{r}^{^{\prime}}%
}^{\ast}(\tau^{^{\prime}})\right]  \right \rangle _{S^{(0)}}  &  =\theta
(\tau-\tau^{^{\prime}})\left \langle a_{\mathbf{r}}(\tau)a_{\mathbf{r}%
^{^{\prime}}}^{\dag}(\tau^{^{\prime}})\right \rangle _{S^{(0)}}\nonumber \\
&  +\theta(\tau-\tau^{^{\prime}})\left \langle a_{\mathbf{r}^{^{\prime}}}%
^{\dag}(\tau^{^{\prime}})a_{\mathbf{r}}(\tau)\right \rangle _{S^{(0)}},
\end{align}
we obtain the relation as%

\begin{align}
&  \left \langle a_{\mathbf{r}}(\tau)a_{\mathbf{r}^{^{\prime}}}^{\ast}%
(\tau^{^{\prime}})\right \rangle _{S^{(0)}}\nonumber \\
&  =\theta(\tau-\tau^{^{\prime}})(\tilde{n}+1)e^{-(-\mu+\tilde{n}U)(\tau
-\tau^{^{\prime}})/\hbar}\nonumber \\
&  +\theta(\tau-\tau^{^{\prime}})\tilde{n}e^{(\mu-(\tilde{n}-1)U)(\tau
^{^{\prime}}-\tau)/\hbar}%
\end{align}
After introducing Matsbara frequencies, the order parameter fields
$\psi_{A\mathbf{k}}(\tau)$ and $\psi_{B\mathbf{k}}(\tau)$ then become%

\begin{align}
\psi_{A\mathbf{k}}(\tau)  &  =\frac{1}{\sqrt{\hbar \beta}}%
%TCIMACRO{\dsum \limits_{\omega_{m}}}%
%BeginExpansion
{\displaystyle \sum \limits_{\omega_{m}}}
%EndExpansion
e^{-i\omega_{m}\tau}\psi_{A\mathbf{k},\omega_{m}},\nonumber \\
\psi_{B\mathbf{k}}(\tau)  &  =\frac{1}{\sqrt{\hbar \beta}}%
%TCIMACRO{\dsum \limits_{\omega_{m}}}%
%BeginExpansion
{\displaystyle \sum \limits_{\omega_{m}}}
%EndExpansion
e^{-i\omega_{m}\tau}\psi_{B\mathbf{k},\omega_{m}}.
\end{align}
At last, we obtain the action near the phase transition point as shown in Eq.
(\ref{eq0g}).

\end{document}